\documentstyle[12pt,aasms4]{article}
\renewcommand{\arraystretch}{0.3}

\begin{document}

\title{Contribution of Type Ia and Type II Supernovae for
Intra-Cluster Medium Enrichment
}

\author{
Shigehiro Nagataki\altaffilmark{1}
 and Katsuhiko Sato\altaffilmark{1,2,3}
}

\noindent
\altaffilmark{1}{Department of Physics, School of Science, the University
of Tokyo, 7-3-1 Hongo, Bunkyoku, Tokyo 113, Japan}\\
\altaffilmark{2}{Research Center for the Early Universe, School of
Science, the University of Tokyo, 7-3-1 Hongo, Bunkyoku, Tokyo 113, Japan} \\
\altaffilmark{3}{
The Institute of Physical and Chemical Research,
Wako, Saitama, 351-01, JAPAN}


\begin{abstract}
The origin of the chemical composition of the intracluster medium
(ICM) is discussed in this paper. In particular, the
contribution from Type Ia supernovae (SNe Ia) to the ICM enrichment is
shown to exist by adopting the fitting formulas which have been used in
the analysis of the solar system abundances.
Our analysis means that we can use the frequency of SNe Ia relative to SNe II
as the better measure than $M_{\rm Fe, SN Ia}/M_{\rm Fe, total}$ for
estimating the contribution of SNe Ia.
Moreover, the chemical compositions of ICMs are shown to be
similar to that of the solar system abundances.
We can also reproduce the sulfur/iron abundance ratio within a factor
of 2, which means that the abundance
problem of sulfur needs not to be emphasized too strongly. We need
more precise observations to conclude whether ICMs really suffer the
shortage problem of sulfur or not.

\end{abstract}


\section{Introduction} \label{intro}
\indent

Primary work on intracluster medium
(ICM) enrichment has focused on iron, since this was the only element
that was accurately measured in a large number of clusters,
prior to the launch of {\it Advanced Satellite for Cosmology and Astrophysics}
($ASCA$, \cite{tanaka94}).
However, $ASCA$ X-ray observations of clusters of galaxies provide
the first opportunity to determine the element abundances of O, Ne,
Mg, Si, S, Ca, Ar, and Fe for the bulk of the intracluster medium
(ICM) in clusters of galaxies (\cite{mushotzky95}).
This abundance ratio pattern in the
ICM provided a unique tool to probe the origin of these
heavy elements.
In particular, a precise analysis of these elemental abundances was
done for four brightest, moderate temperature clusters of galaxies (Abell
496, 1060, 2199, and AWM 7).
The mean abundances of O, Ne, Si, S and Fe were reported as
0.48, 0.62, 0.65, 0.25, and 0.32, respectively, relative to
the solar photospheric abundances (\cite{mushotzky96}).

Loewenstein \& Mushotzky (1996) tried to explain the observed
abundance ratios of these elements with respect to Fe (since Fe is the
most accurately measured element) by the calculated
yields from Type II supernovae (SNe II). They concluded that the
abundance ratio pattern of ICM is very similar to that of SNe II.
However, they emphasized that a significant SNe Ia
contribution to the Fe enrichment could not be ruled out.
In fact, Ishimaru \& Arimoto (1997) concluded that a SNe Ia iron
contribution of $50 \%$ or higher to the ICM enrichment could not be
ruled out and might be favored by the observations.
In the end, Gibson et al. (1997) pointed out the large uncertainties
of the adopted massive star physics and uncertainties in the resulting
SNe II's chemical composition, which make it impossible to
determine the contribution of SNe Ia to the iron enrichment of ICM.
As a result, we have no way of estimating the
contribution of SNe Ia to the ICMs.

There is another problem pointed out by Mushotzky et al. (1996),
Loewenstein \& Mushotzky (1996), and Gibson et al. (1997).
It is the lack of sulfur in the ICMs.
Gibson et al. (1997) insist that there would appear to be no escape
from the fact that the sulfur abundances are at odds with a SNe
II-dominated origin to the ICM iron abundance (although the
uncertainties are large).
This appears to be entirely independent of adopted SNe II yields
(\cite{gibson97}).
Quantitatively, the observed ratio S/Fe favors the ICM
SNe Ia iron fractional contribution to be larger than $ 80 \%$.
On the other hand,
O/Fe, Si/Fe, Mg/Fe, and Ne/Fe favor a ratio under $60 \%$.

In this paper, we adopt the fitting formula used in the analysis of
the solar system abundances (\cite{tsujimoto95}) and a $\chi ^2$ fitting
formula to obtain the relative contributions of SNe Ia and SNe II to
the enrichment of ICM. We anticipate that these formulas will become
good measures for estimating the fractional contribution of SNe Ia.
We will also adopt many models of SNe Ia (model W7 and WDD2,
\cite{nomoto97}) and SNe II (\cite{nagataki97})
which take into account the effect of axisymmetric explosion, in
order to reexamine the shortage problem of sulfur.

In section~\ref{observation}, the results of X-ray
observations of clusters of galaxies are presented.
In section~\ref{model}, we show a range of 
calculated yields from SNe Ia and SNe II.
In section~\ref{results}, we present our analysis and results. Summary
and discussion are given in section~\ref{summary}.

\section{ Observations } \label{observation}
\indent

We show the abundances derived from SIS (solid state
imaging spectrometer) in Table~\ref{obs}. [$\rm M_i/Fe$] is defined as 
$\log(\rm M_i/Fe)_{\rm ICM} - \log(\rm M_i/Fe)_{\rm solar}$.
The error bars indicate $90 \%$ confidence intervals.
We note that the meteoritic
abundance scale of Anders $\&$ Grevesse (1989) is adopted.
Oxygen, silicon, and iron have the best-determined abundances. On the
other hand, the accuracy for magnesium, argon, and calcium is very
poor (\cite{mushotzky96}).

\placetable{obs}

\section {Models of SNe Ia and SNe II \label{model}}
\indent

There is a range of SNe II models available in the literature. In the
present situation, there are large differences in the predicted yields 
since there is a
wide variety of input physics (e.g., criterion for convection,
reaction rates, and the way to initiate the shock wave)
(\cite{aufderheide91}; \cite{gibson97}).

In order to compare the differences quantitatively, we make use of the 
formalism presented by Ishimaru $\&$ Arimoto (1997). The $i$th element
yield averaged over the SNe II progenitor initial mass function (IMF), 
$\phi \propto m^{-x}$, is written as
\begin{eqnarray*}
<y_{\rm i,SNII}> = \frac{\int^{m_u}_{m_l}y_{\rm i,SNII}(m) \phi (m)
m^{\rm -1} \rm d \it m}{\int ^{m_u}_{m_l} \phi (m) m^{\rm -1} \rm d
\it m}.
\end{eqnarray*}
The bounds for SNe II progenitors are taken to be $m_l = 10 M_{\odot}$ 
and $m_u = 50 M_{\odot}$, respectively, and an IMF slope $x = 1.35$
(\cite{salpeter55}) is adopted throughout. While this slope is of
prime importance for arguments concerning the absolute mass of
elements in the ICM (e.g., \cite{loewenstein96}), it is less so for
abundance ratios (\cite{gibson97}).

In imitation of Gibson et al. (1997), we summarize $<y_{\rm i,SNII}>$
for a range of SNe II models in Table~\ref{sneii}. We add the yields
predicted by Nagataki et al. (1997) and explain it here.
They performed 2-dimensional hydrodynamical calculations and studied 
the changes of the chemical compositions using a large nuclear
reaction network containing 242 nuclear species.
Model N97S1 in Table~\ref{sneii} is a spherical explosion model, N97A1 and
N97A3 are models of
axisymmetric explosions in an increasing order of the degree of
deviation from spherical symmetry. 
For the axisymmetric models N97A1 and N97A3, the initial velocity
behind the shock
wave is assumed to be radial and proportional to $r \times 
\frac{1 + \alpha \cos (2 \theta)}{1 + \alpha }$, where $r$, $ \theta $, 
and $ \alpha $ are the radius, the zenith angle and the model
parameter that determines the degree of deviation from spherical
symmetry, respectively.
Nagataki et al. (1997) took $ \alpha = 0 $ for model N97S1, $ \alpha =
\frac{1}{3} $ for model N97A1, and
$ \alpha = \frac{7}{9} $ for model N97A3. The larger $ \alpha $ gets, the more
asymmetric the explosion becomes. They used the same distribution
also for the thermal energy.
Half of the total energy appears as kinetic energy and the other 
half as thermal energy.

As we can see from Table~\ref{sneii}, almost all $<y_{\rm i,SNII}>$ 
of Nagataki et al. (1997) are in the range of uncertainties of
theoretical predictions. However, the amount of sulfur tends to become
smaller as $\alpha$ gets larger. In fact, N97A3 predicts the least
amount of sulfur among all models. This tendency is expected to be
good for the reproduction of the observed amount of sulfur in ICMs.

We also show the
SNe Ia yields $<y_{\rm i,SNe Ia}>$ at the lower portion of
Table~\ref{sneii}.
W7 is the model of the simple
deflagration and WDD2 is that of the delayed detonation (\cite{nomoto97}).
As pointed by Gibson et al. (1997), all SNe Ia yields have been tied
exclusively to the W7 model. In this paper, the uncertainty of SNe Ia
yields is taken into consideration for the first time.

\placetable{sneii}

\section{ Analysis $\&$ Results} \label{results}

\subsection{Previous Analysis}\label{analysis1}
\indent

\begin{figure}
\epsscale{1.0}
\plottwo{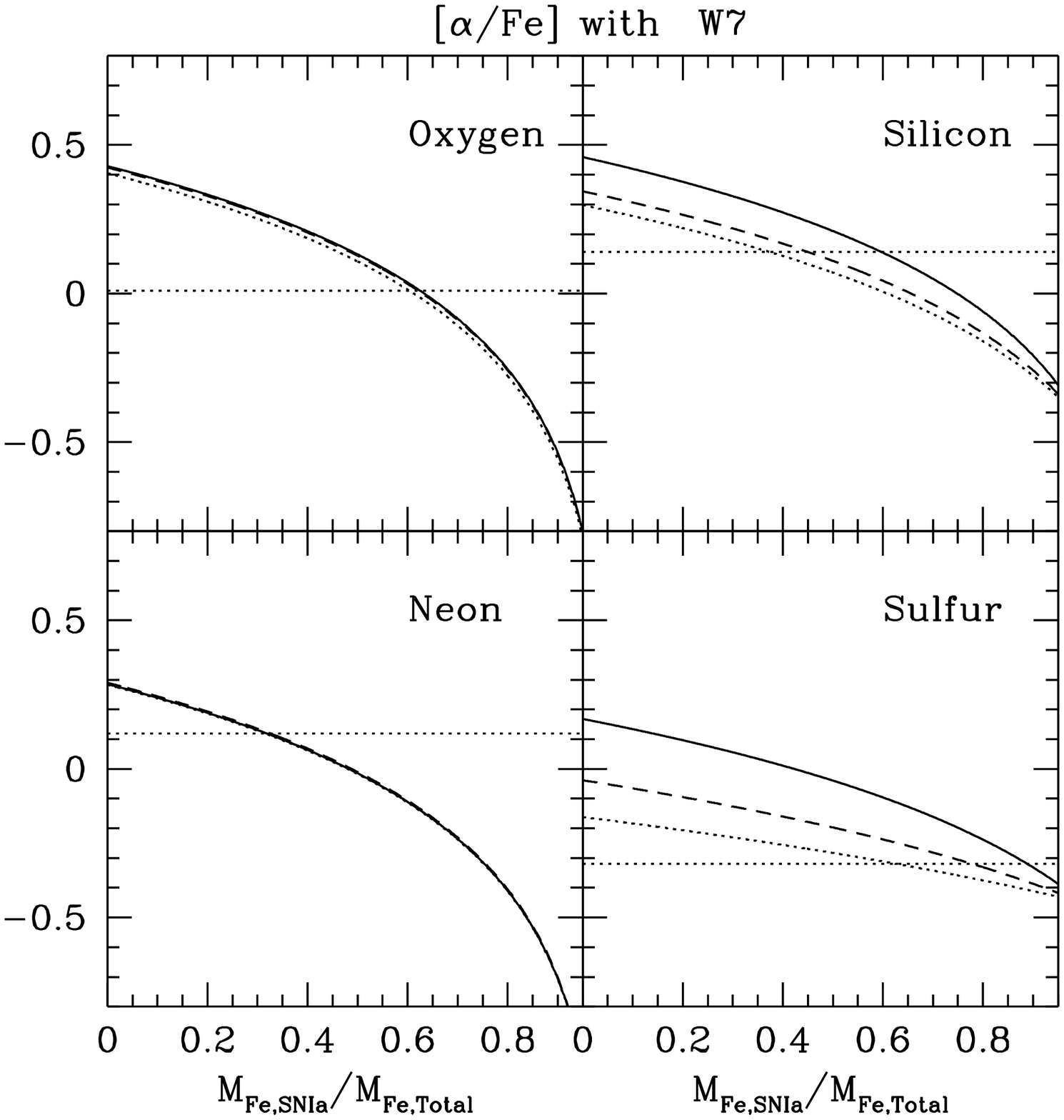}{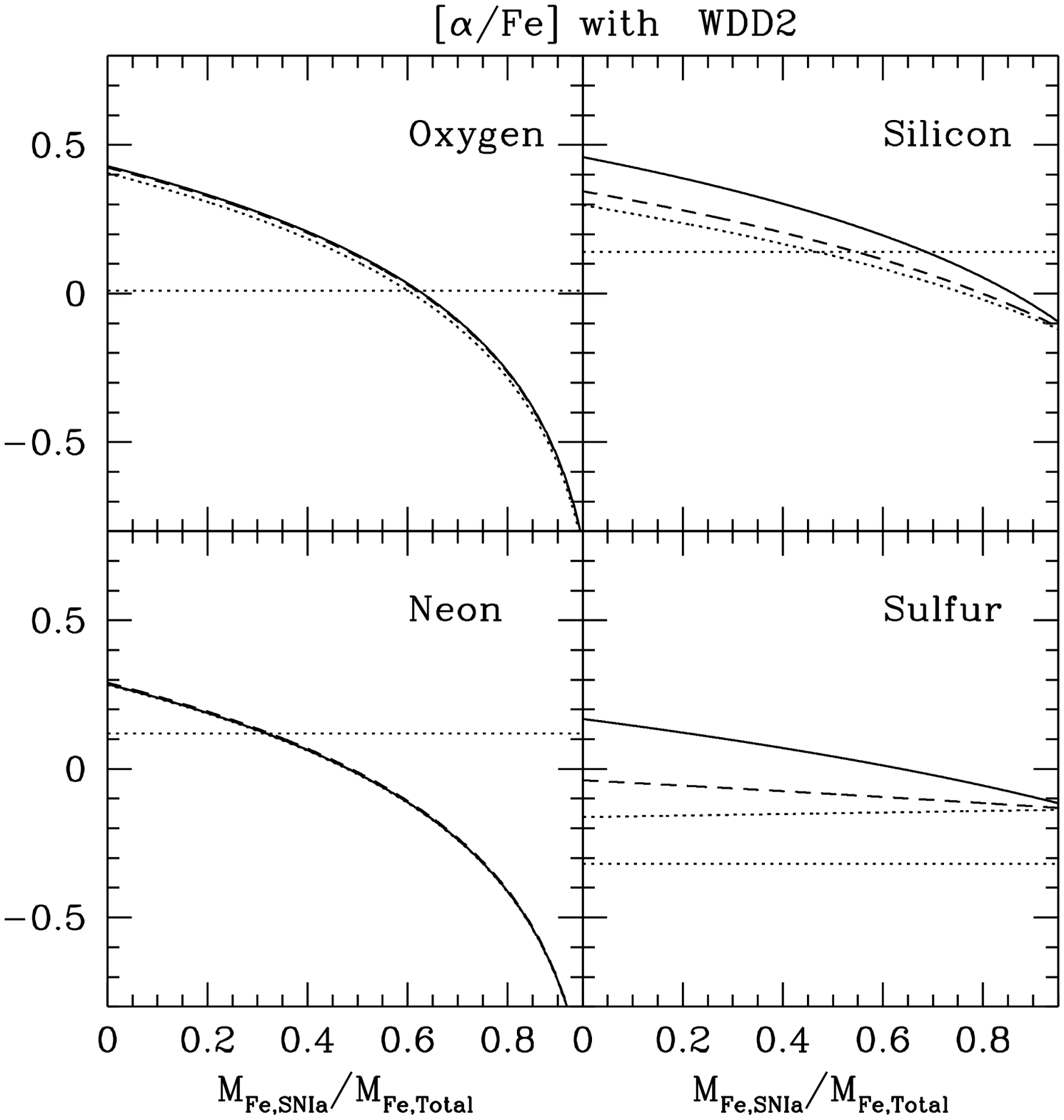}
\figcaption{
Theoretical ratio of relative abundances of heavy elements,
[O/Fe], [Ne/Fe], [Si/Fe], and [S/Fe] as a function of the SN Ia
fraction in the iron synthesis. Solid, short-dashed, and dot lines
correspond to N97S1, N97A1, and N97A3 models, respectively. Left: W7 model is
adopted for SNe Ia. Right: WDD2 model is adopted.
\label{fig1}}
\end{figure}

At first, the analysis which is done in the previous papers is reexamined.
We calculate the abundances ratios [$\rm M_i/Fe$] for $\rm
M_i$ = O, Ne, Si, and S as a function of $M_{\rm Fe, SN Ia}/M_{\rm 
Fe, total}$. $M_{\rm Fe, SN Ia}/M_{\rm Fe, total}$ is the
contribution of SNe Ia to the ICM iron enrichment and is given as
\begin{eqnarray}
\frac{M_{\rm Fe, SN Ia}}{M_{\rm Fe, total}} = \frac{r <y_{\rm
Fe,SNIa}>}{r <y_{\rm Fe,SNIa}>+(1- r)<y_{\rm Fe,SNII}>}, \;\;\; (0 \le
r \le 1)
\end{eqnarray}
where $r$ indicates the relative occurrence frequency of SNe Ia.
The result is shown in Figure~\ref{fig1}.
Horizontal dot lines mean the average of $ASCA$ SIS data.

\placefigure{fig1}

[O/Fe] and [Ne/Fe] are not changed even when the model of SNe II is changed.
This is because the abundances of O and Ne are not determined mainly by
the explosive nucleosynthesis but by the nucleosynthesis during the
stellar evolution.
We also note that the abundance of Fe is mainly determined by the amount of
$\rm ^{56}Ni$, which decays to $\rm ^{56}Fe$. Since the position of
the mass cut is also determined by the amount of $\rm ^{56}Ni$, Fe
abundance is not changed among three models, too (\cite{nagataki97}).
On the other hand, [Si/Fe] and [S/Fe] are sensitive to the selection
of the model of SNe II. This is because Si and S are mainly
synthesized during the explosive nucleosynthesis. Si and S abundances
tend to decrease along with the degree of deviation from spherical
symmetry. Because of
this, the [S/Fe] ratios give smaller $M_{\rm Fe, SNIa}/M_{\rm Fe, 
Total}$ when W7 and the axisymmetric models are adopted.

As a result, the observations of O, Ne, Si, S, and Fe favor the
ICM SNe Ia iron fractional contribution in the range $30 \% \le
M_{\rm Fe, SNIa}/M_{\rm Fe, Total} \le 60 \%$ for the models W7 and N97A3.
On the other hand, it is in the range $30 \% \le M_{\rm Fe,
SNIa}/M_{\rm Fe, Total} \le 90 \%$ for the models W7 and N97S1.
Even worse, the averaged [S/Fe] ratio can not be reproduced
when model WDD2 is adopted, although the theoretical [S/Fe] ratio with
a proper ratio of $M_{\rm Fe, SNIa}/M_{\rm Fe, Total}$ can be within
the error bars.

As we see, the ratio of $M_{\rm Fe, SNIa}/M_{\rm Fe, Total}$ is very hard 
to determine, which was also indicated in the previous papers
(\cite{loewenstein96}; \cite{ishimaru97}; \cite{gibson97}).

\subsection{g($\zeta$) and $\chi^2$ -- fitting}\label{analysis2}
\indent

Next, we define $y_{\rm SNIa}$ as the sum of $<y_{i,\rm SNIa}>$, which is the
heavy element mass of SNe Ia. $y_{\rm SNII}$ is defined in the same way.
We also define the abundance pattern $x_i$ as
\begin{eqnarray}
x_i = \zeta y_{i,\rm SNIa}/y_{\rm SNIa} + (1- \zeta)y_{i,\rm
SNII}/y_{\rm SNII}  \;\;\; (0 \le \zeta \le 1)
\end{eqnarray}
which is to be compared with the observed $x_{i,\rm ICM}$.
$x_{i,\rm ICM}$ is defined as $Z_i/\sum_i Z_i$, where $Z_i$ is
the observed abundance of the $i$-th element per unit mass.
$\zeta$ is the mass fraction of SNe Ia's matter in ICM. The relation
between $r$ and $\zeta$ is written as
\begin{eqnarray}
\frac{r}{1-r} = \frac{\omega_{\rm II}}{\omega_{\rm Ia}}
\frac{y_{\rm II}}{y_{\rm Ia}}\frac{\zeta}{(1-\zeta)}
\end{eqnarray}
where $\omega_{\rm Ia}$ and $\omega_{\rm II}$ represent the mass fraction of
heavy elements ejected into the interstellar gas from SNe Ia and SNe
II, respectively. These values are estimated to be 0.27 and 0.22 in
the solar neighborhood from the numerical calculation (\cite{tsujimoto95}).

The most probable value of $\zeta$ = $\zeta_p$ is
determined by minimizing the following function (Yanagida et al.\ 1990):
\begin{eqnarray}
g(\zeta) = \sum_{i=1}^{n} [\log x_{i,\rm ICM} - \log x_i]^2/n
\end{eqnarray}
where $i$ runs over the heavy elements considered in
the minimization procedure. We summarize those elements in Table~\ref{obs}.
We will give another method to obtain $\zeta_p$, that is, $\chi ^2$ fitting. 
As is well known,
\begin{eqnarray}
\chi ^2 = \sum_{i=1}^{n} \left( \frac{x_i(\zeta) - x_{i,\rm
ICM}}{\sigma_{i}} \right)^2
\end{eqnarray}
obeys the $\chi^2$ distribution for $n$ degrees of freedom.
We note that $\sigma_{i}$ is the standard deviation for $x_{i, \rm ICM}$.
Since we have no information about $\sigma_{i}$, we use the error bars 
which indicate 90$\%$ confidence intervals (\cite{ishimaru97}) instead.

\begin{figure}
\epsscale{1.0}
\plottwo{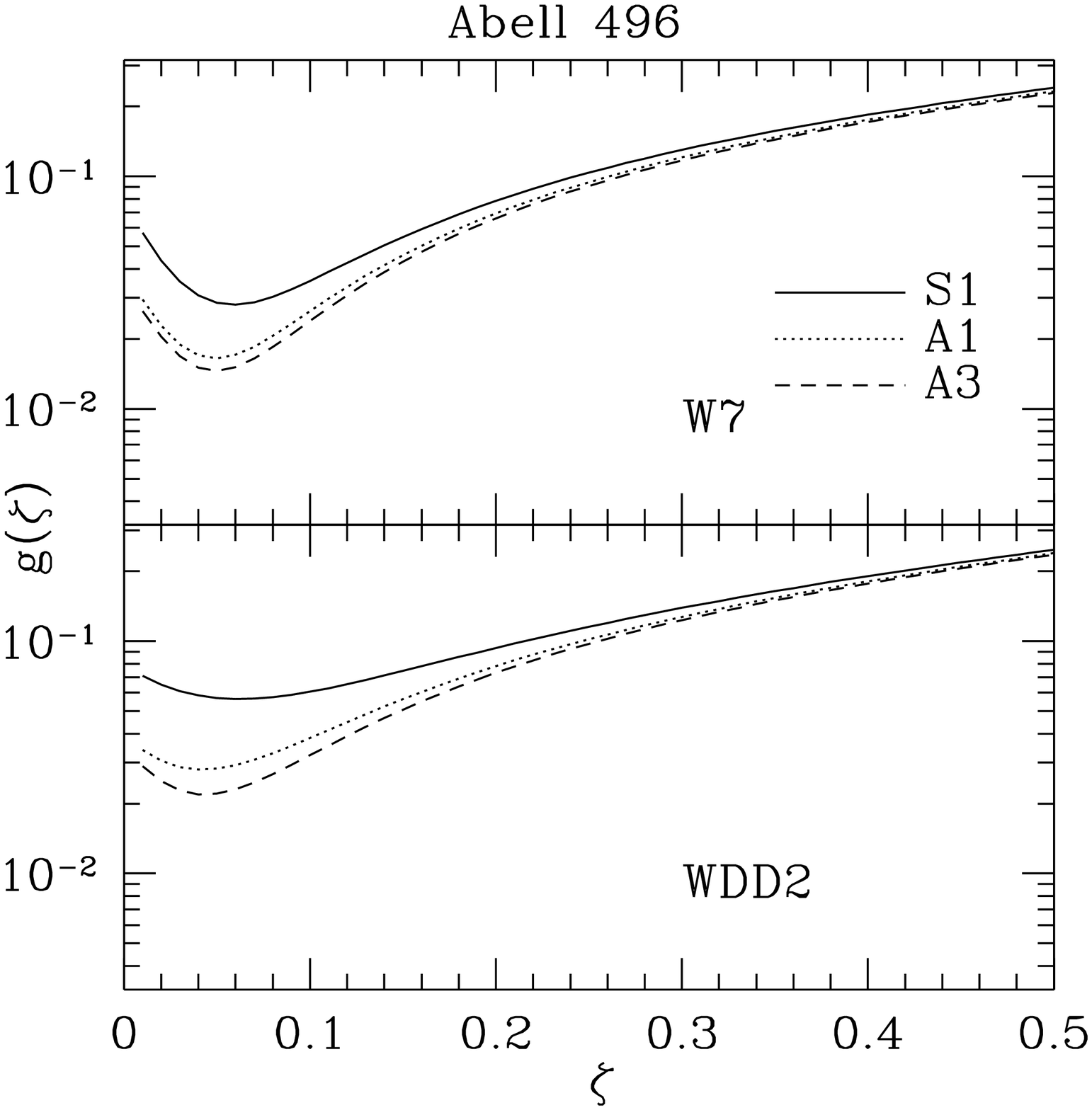}{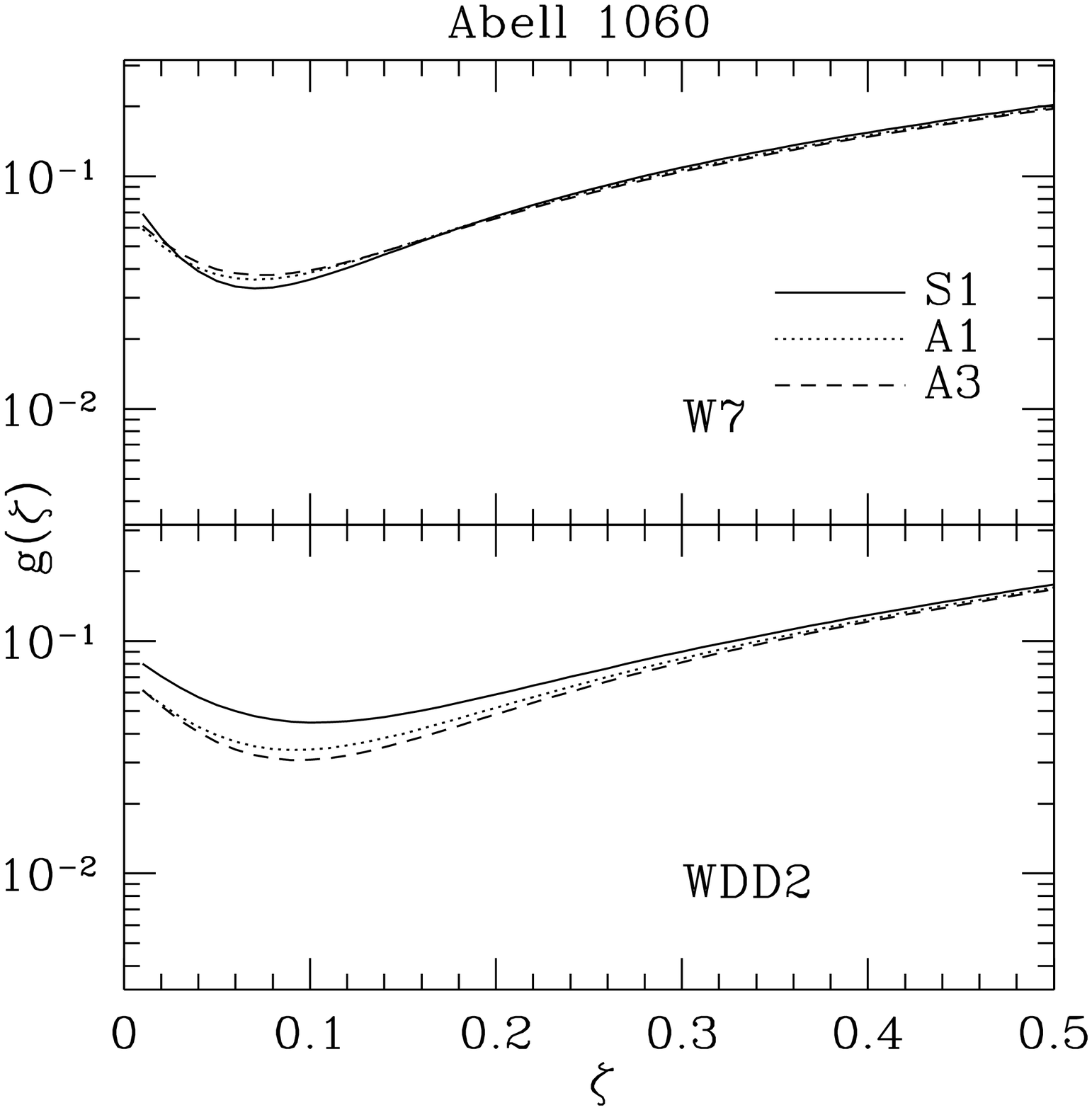}
\figcaption{
$g(\zeta)$ as a function of $\zeta$. W7 and WDD2 are used for SNe Ia.
Left: Abell 496. Right: Abell 1060.
\label{fig2}}
\end{figure}

\begin{figure}
\epsscale{1.0}
\plottwo{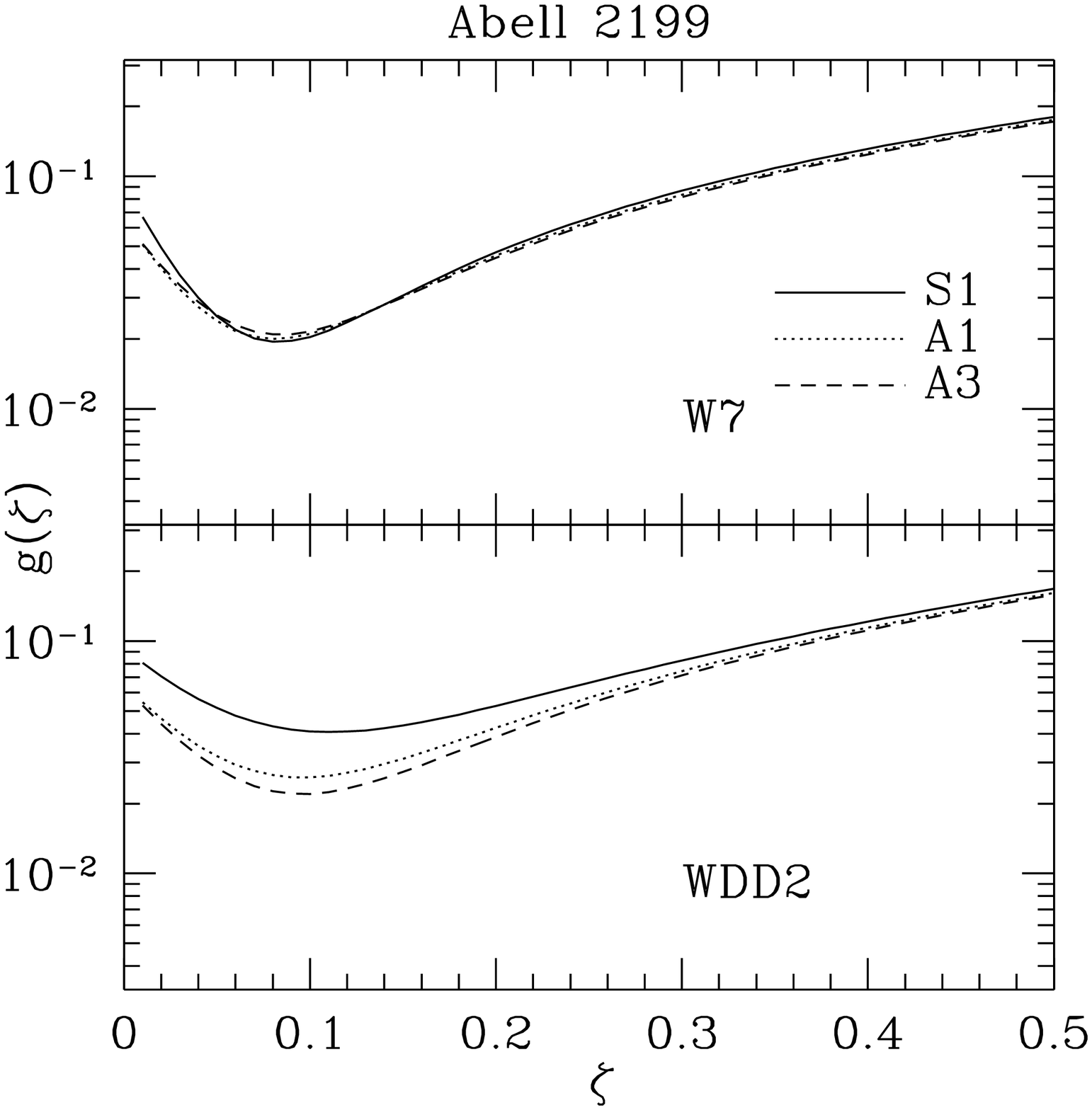}{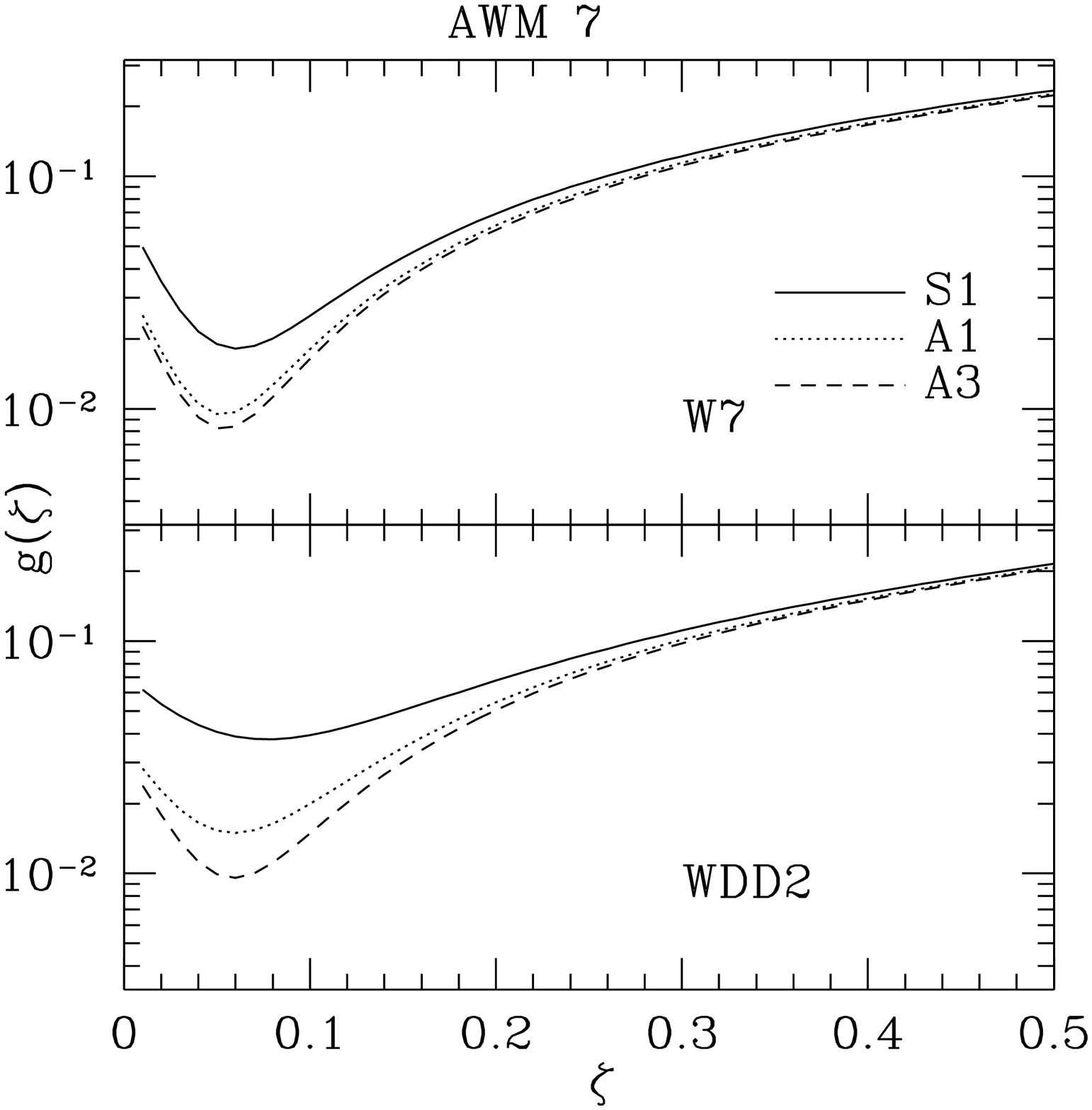}
\figcaption{
Same as figure 2 but for Abell 2199 and AWM 7.
\label{fig3}}
\end{figure}

\begin{figure}
\epsscale{1.0}
\plottwo{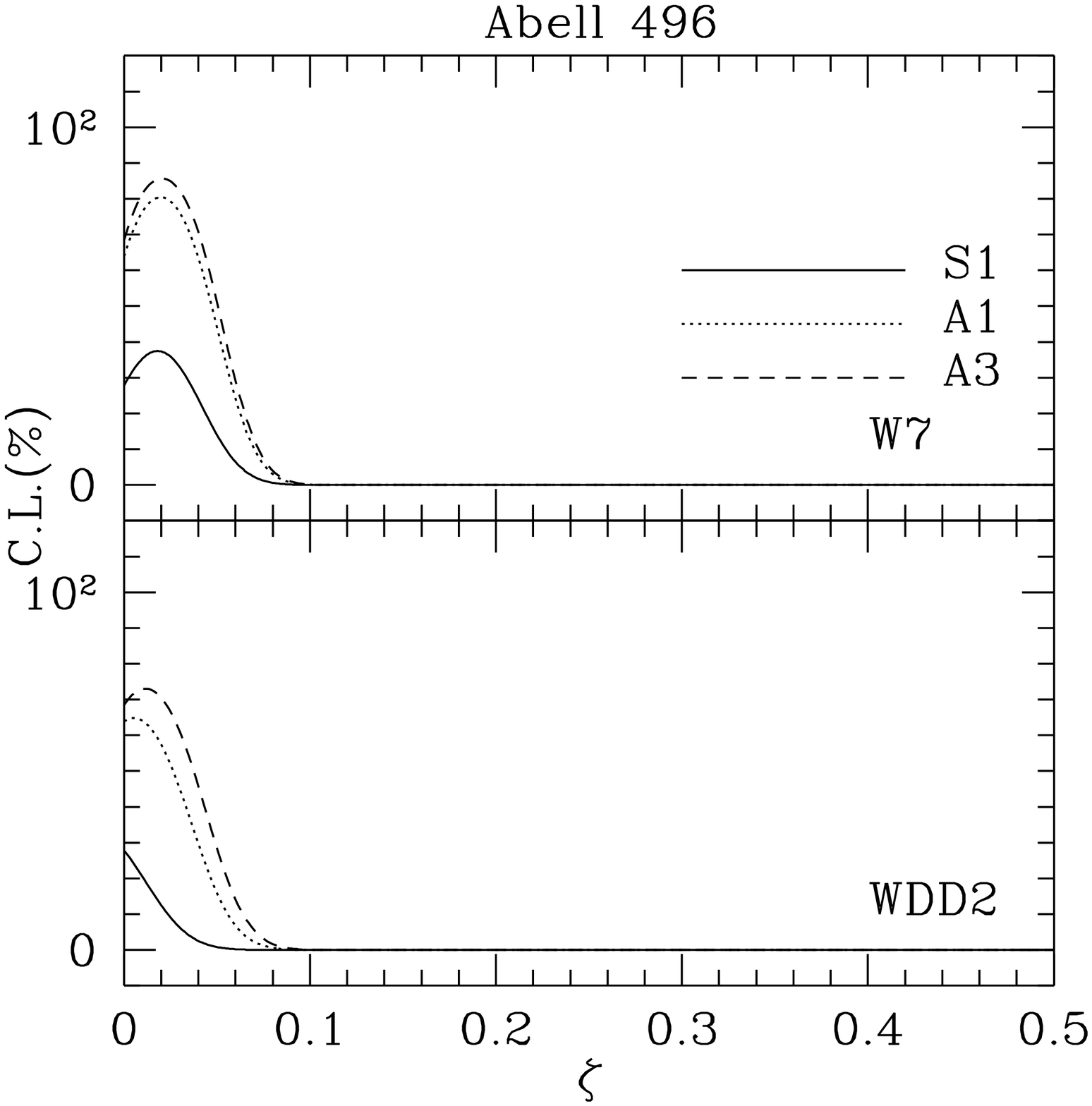}{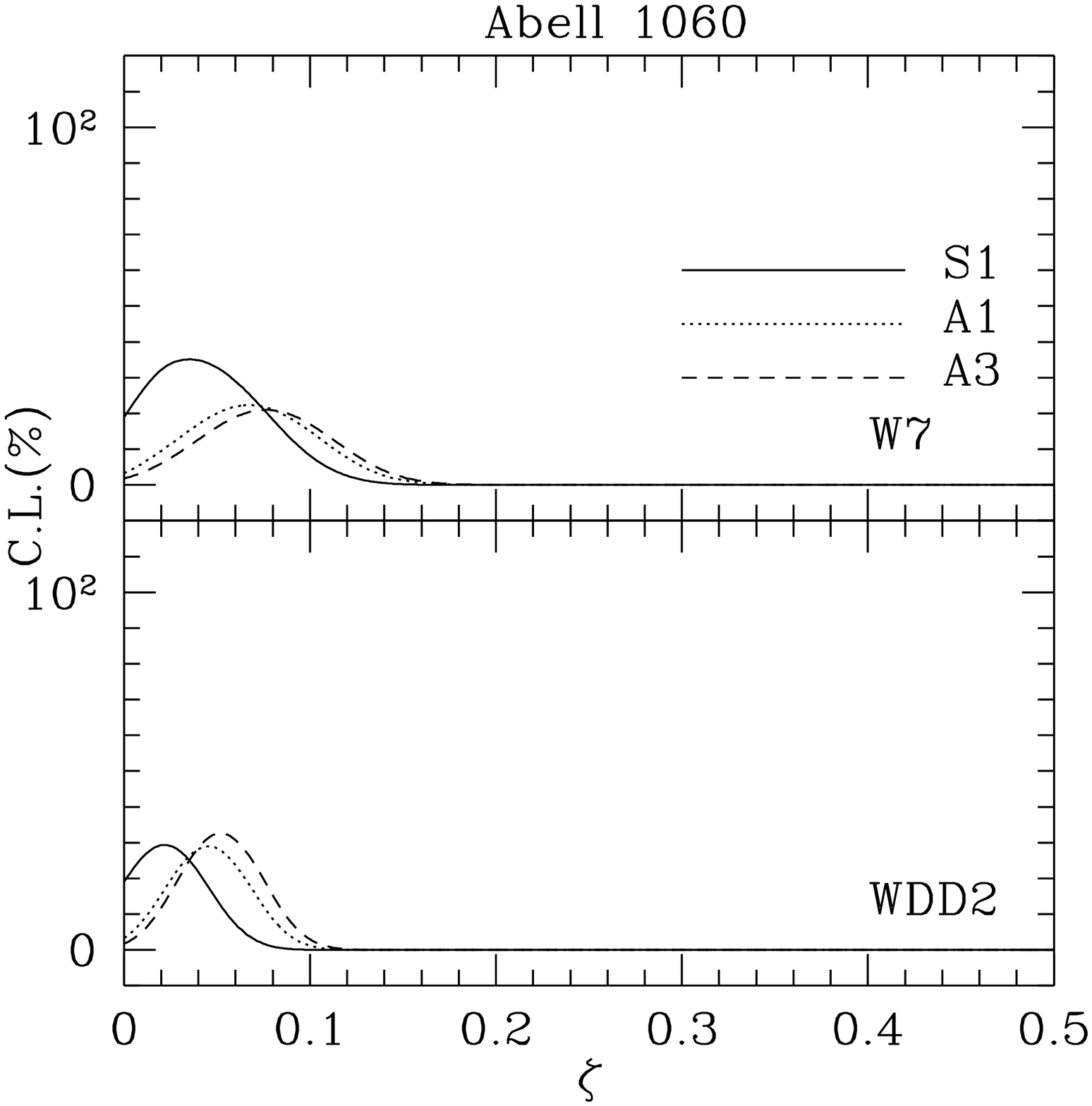}
\figcaption{
Confidence level ($\%$)of $\chi ^2$ fitting as a function of $\zeta$.
Left: Abell 496. Right: Abell 1060.
\label{fig4}}
\end{figure}

\begin{figure}
\epsscale{1.0}
\plottwo{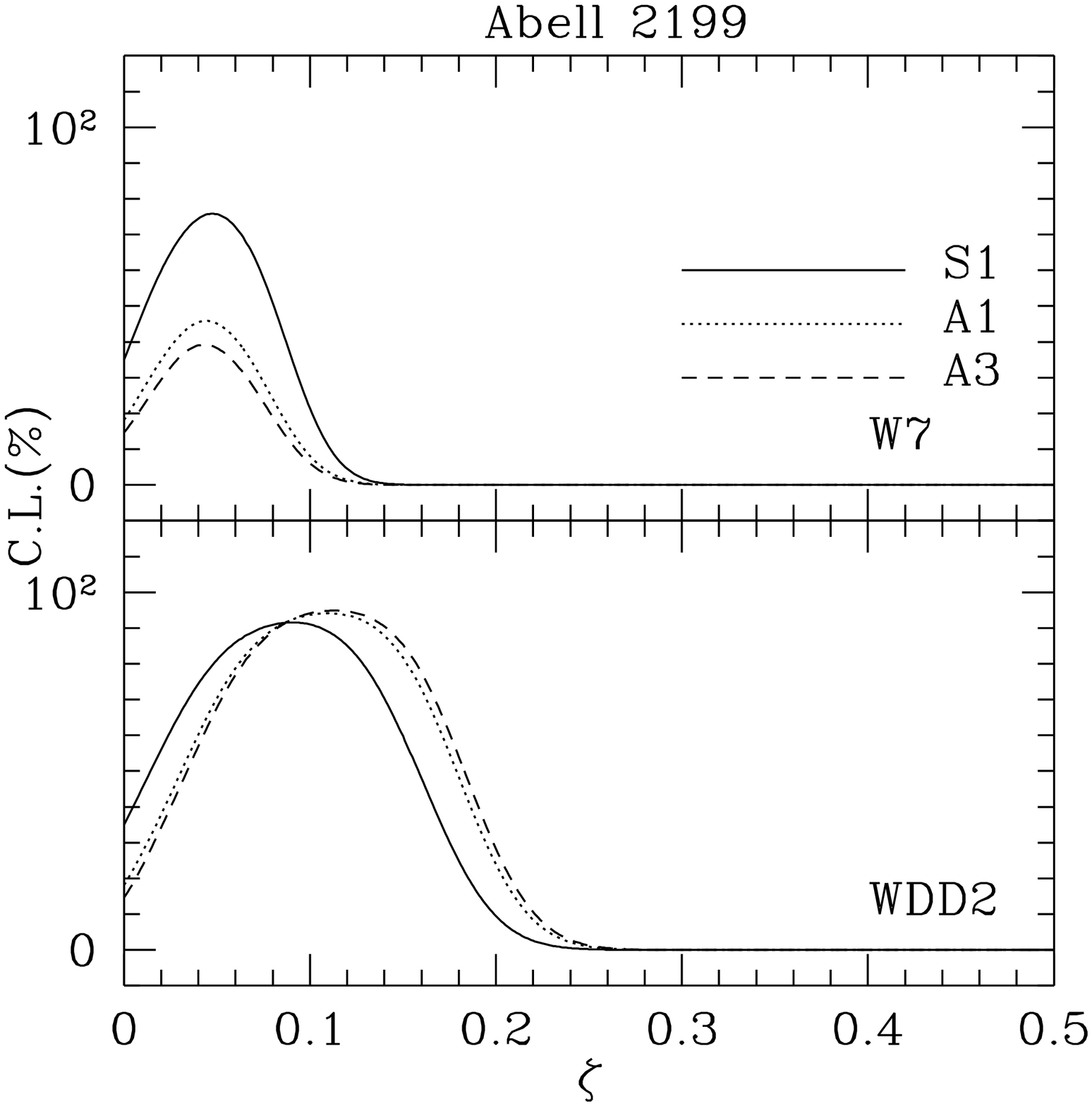}{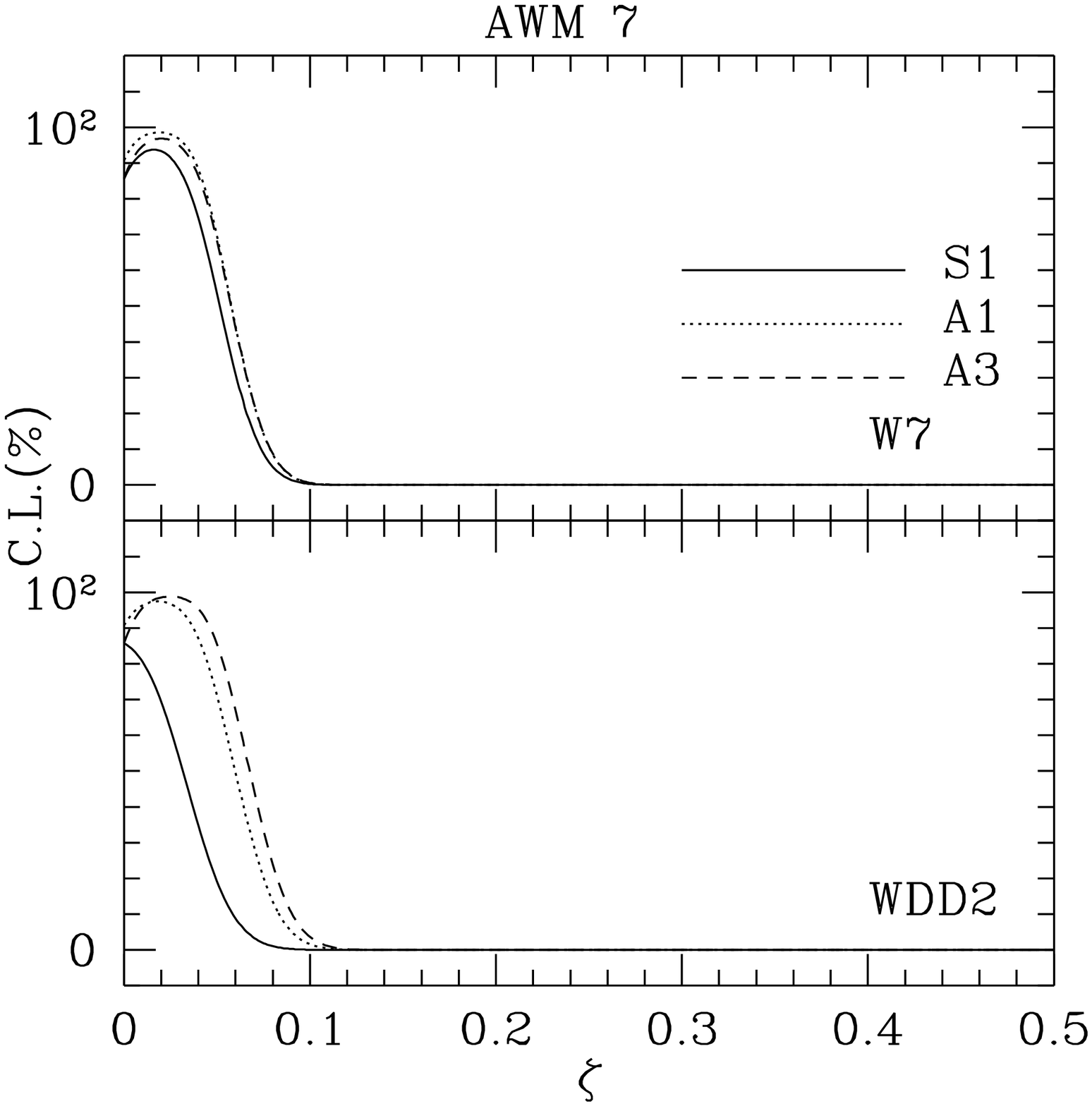}
\figcaption{
Same as figure 4 but for Abell 2199 and AWM 7.
\label{fig5}}
\end{figure}

We show the results for $g(\zeta)$ and $\chi ^2$ in Figure ~\ref{fig2}, 
~\ref{fig3}, ~\ref{fig4}, and ~\ref{fig5}.
It should be noted that $g(\zeta)$ and $\chi ^2$ have a local
minimum/maximum for
each model, which means there is a favorable value of $\zeta$ for the
reproduction of the chemical composition of ICMs.
Moreover, $\zeta_p$ is in the range of $0 < \zeta_p  < 0.1$ for almost all
clusters, which represents that the contributions of SNe Ia to the ICM
enrichment are similar among four ICMs.
It should also be noted that $\zeta_p = 0.09 \pm 0.01$ is obtained for 
the solar system abundances (\cite{tsujimoto95}), which will mean
that the chemical compositions of ICMs are similar to that of the
solar system abundances.

\begin{figure}
\epsscale{1.0}
\plotone{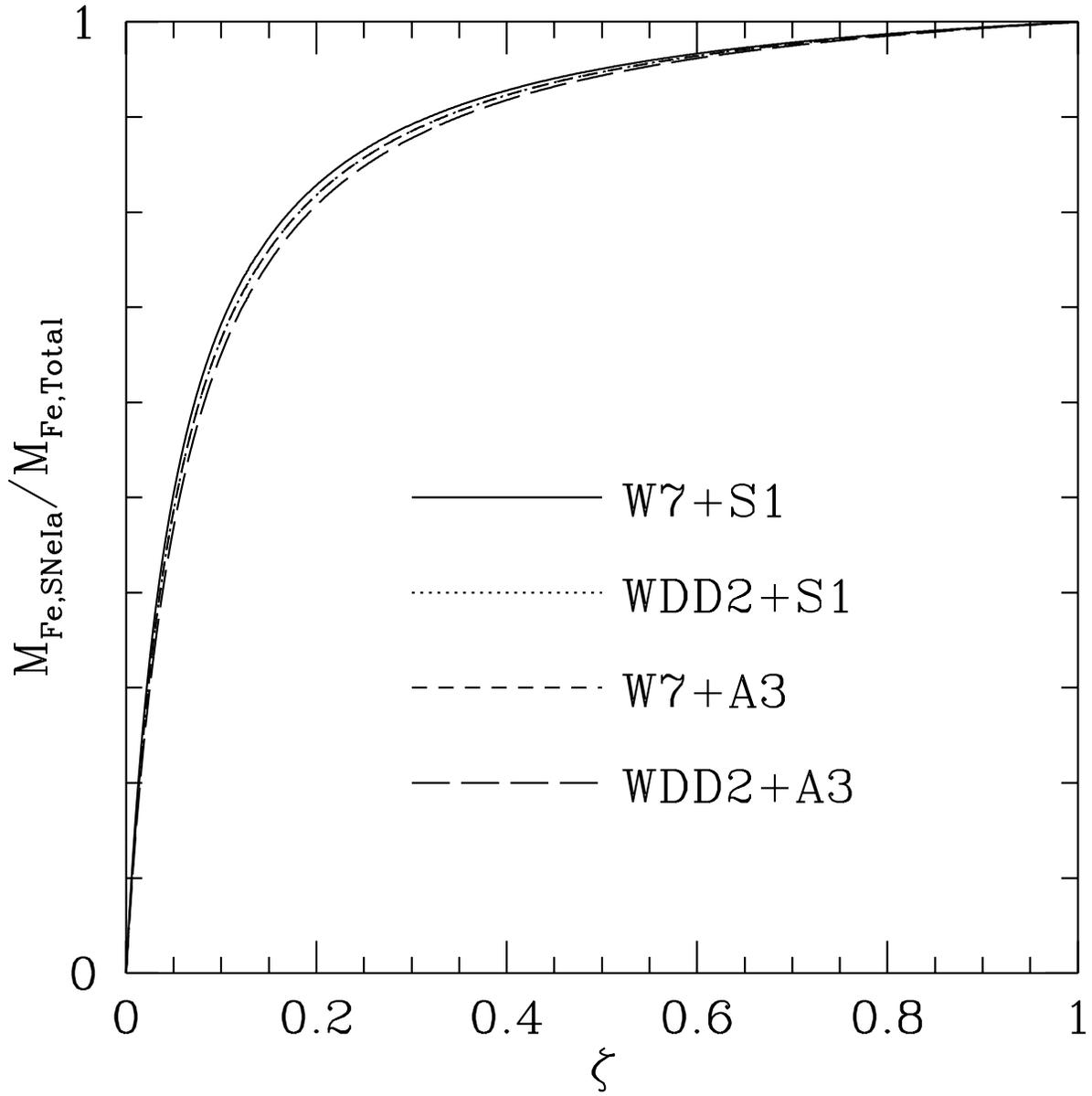}
\figcaption{
Relation between $\zeta$ and $M_{\rm Fe, SNe Ia}/M_{\rm Fe,Total}$.
Solid, dot, short-dashed, and long-dashed lines correspond to W7+N97S1,
WDD2+N97S1, W7+N97A3, and WDD2+N97A3, respectively.
\label{fig6}}
\end{figure}

Now, there is one question: why can $\zeta_p$ be determined in the narrow
range $\zeta_p \le 0.1$ in spite of the large uncertainty of the
inferred ratio
$M_{\rm Fe, SN Ia}/M_{\rm Fe, total}$ ? The answer is shown in
Figure~\ref{fig6}, which gives the relation between $\zeta$ and $M_{\rm
Fe, SN Ia}/M_{\rm Fe, total}$. As can be seen from the figure, 
$M_{\rm Fe, SN Ia}/M_{\rm Fe, total}$ is very sensitive to $\zeta$
if $\zeta$ is in the range of $0 \leq \zeta_p \leq 0.1$. This is
because the mass fraction of Fe in the SNe Ia's ejecta is quite high.
As a result, it is impossible to determine $M_{\rm Fe, SN
Ia}/M_{\rm Fe, total}$ exactly, as we showed in
section~\ref{analysis1}. On the other hand, we can use $\zeta$ as a 
better measure than $M_{\rm Fe, SN Ia}/M_{\rm Fe, total}$ to represent 
the contribution of SNe Ia to the ICM enrichment.

\placefigure{fig2}
\placefigure{fig3}
\placefigure{fig4}
\placefigure{fig5}
\placefigure{fig6}

\begin{figure}
\epsscale{1.0}
\plottwo{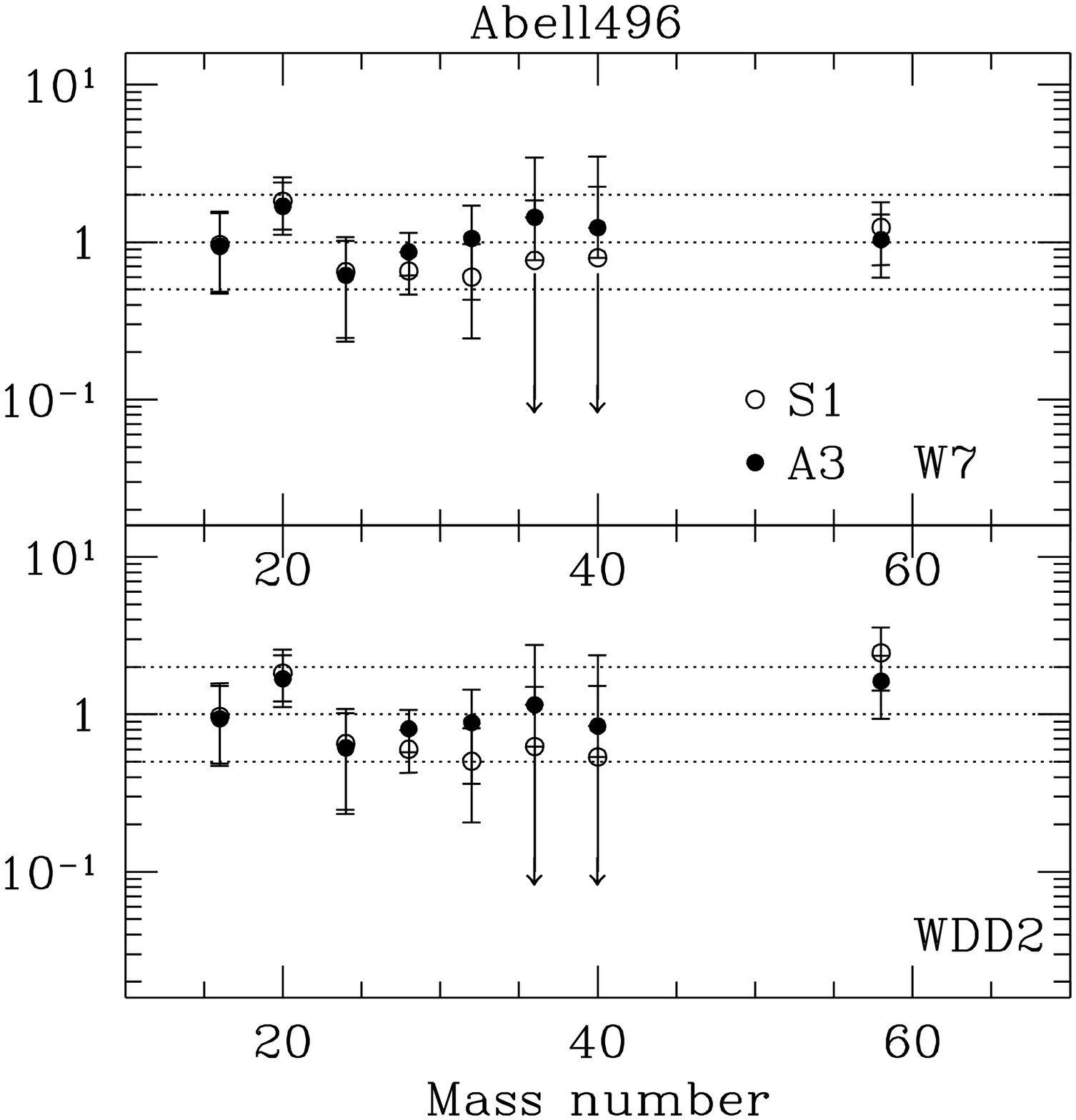}{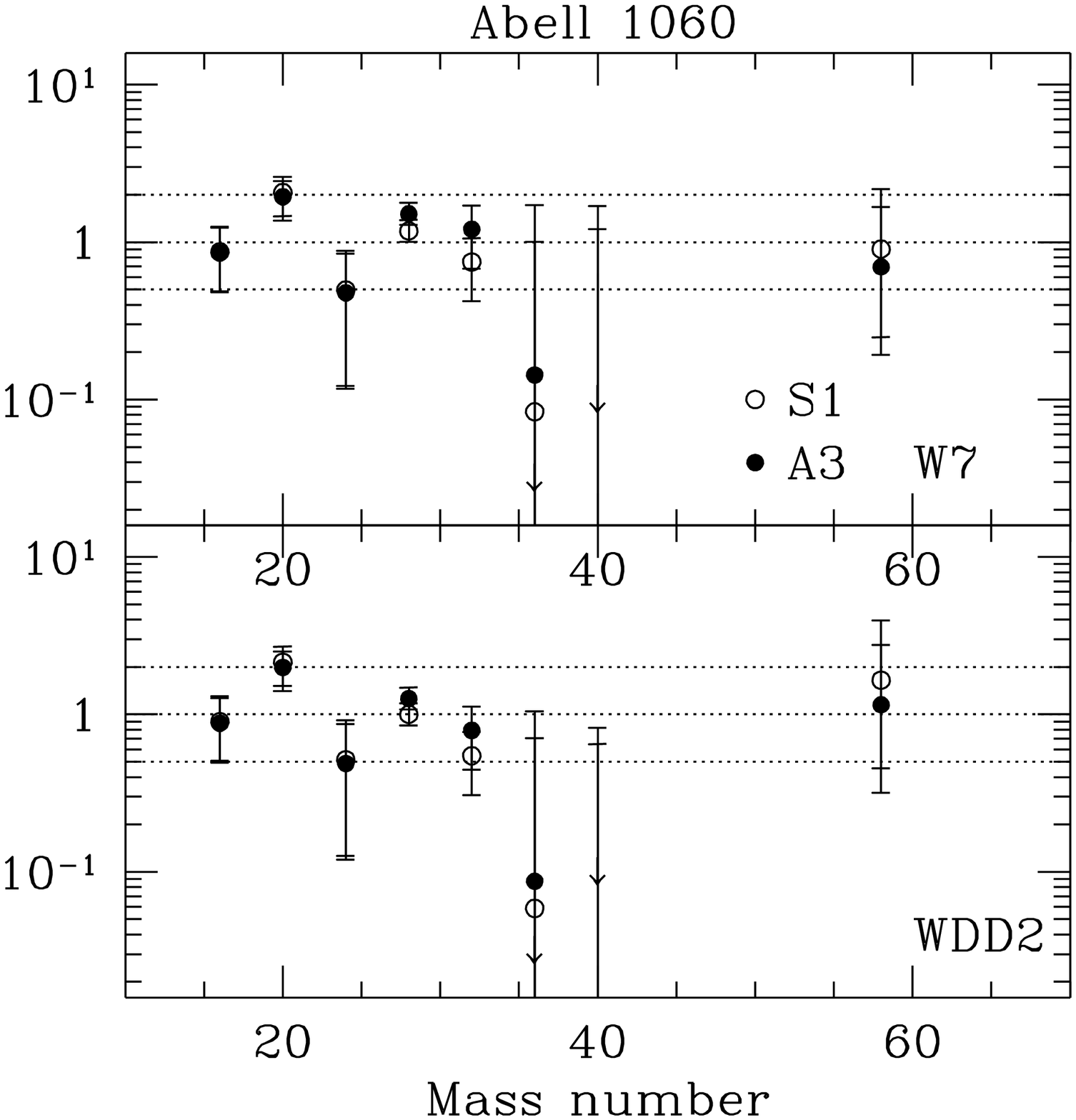}
\figcaption{
Normalized abundance pattern $(x_i/x_{\rm Fe})_{\rm ICM} / (\it
x_i/x_{\rm Fe})_{\rm SN}$ with the most probable value $\zeta = \zeta_p$ which
is obtained from the analysis of $g(\zeta)$. Left: Abell 496. Right:
Abell 1060.
\label{fig7}}
\end{figure}

\begin{figure}
\epsscale{1.0}
\plottwo{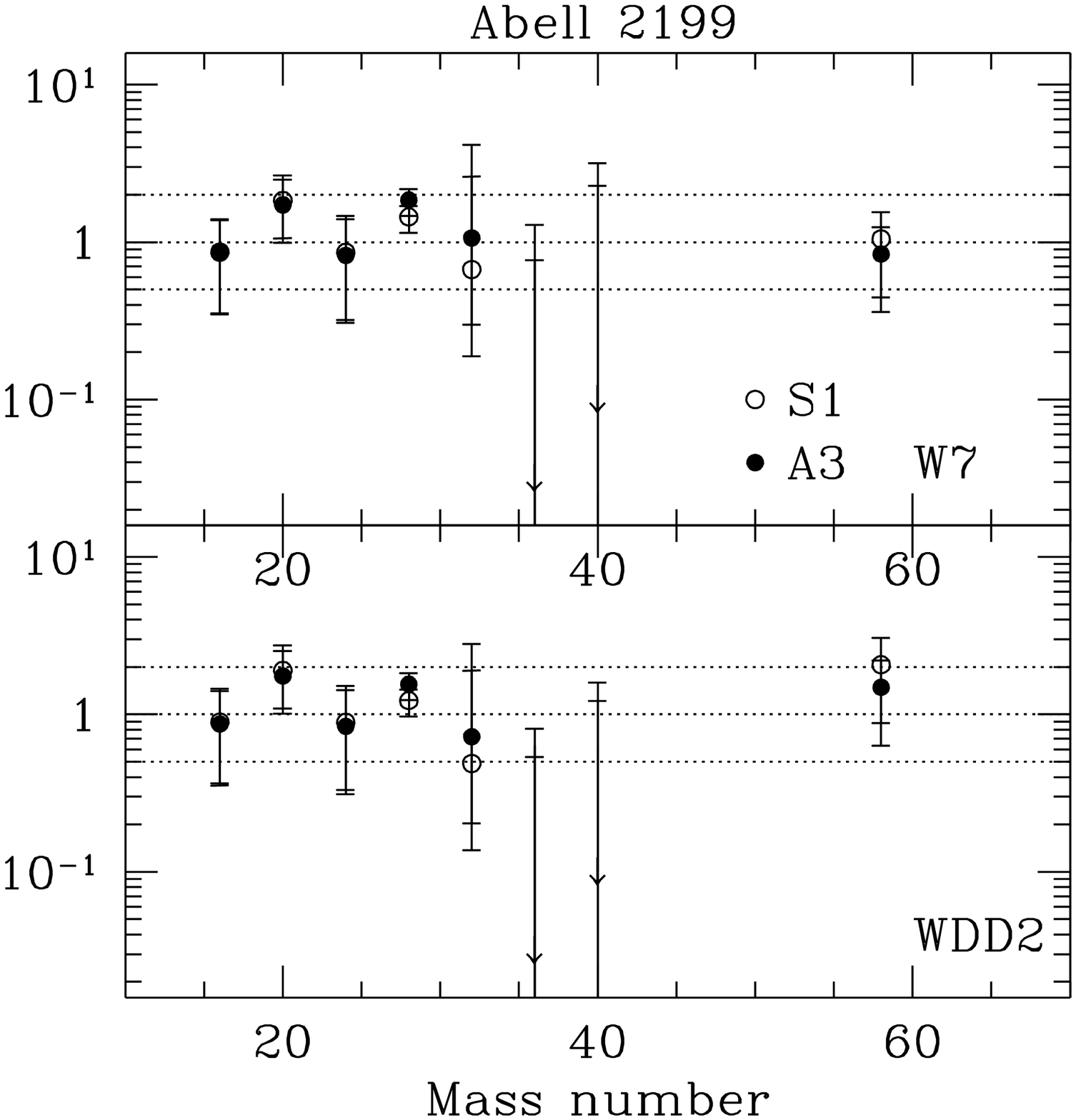}{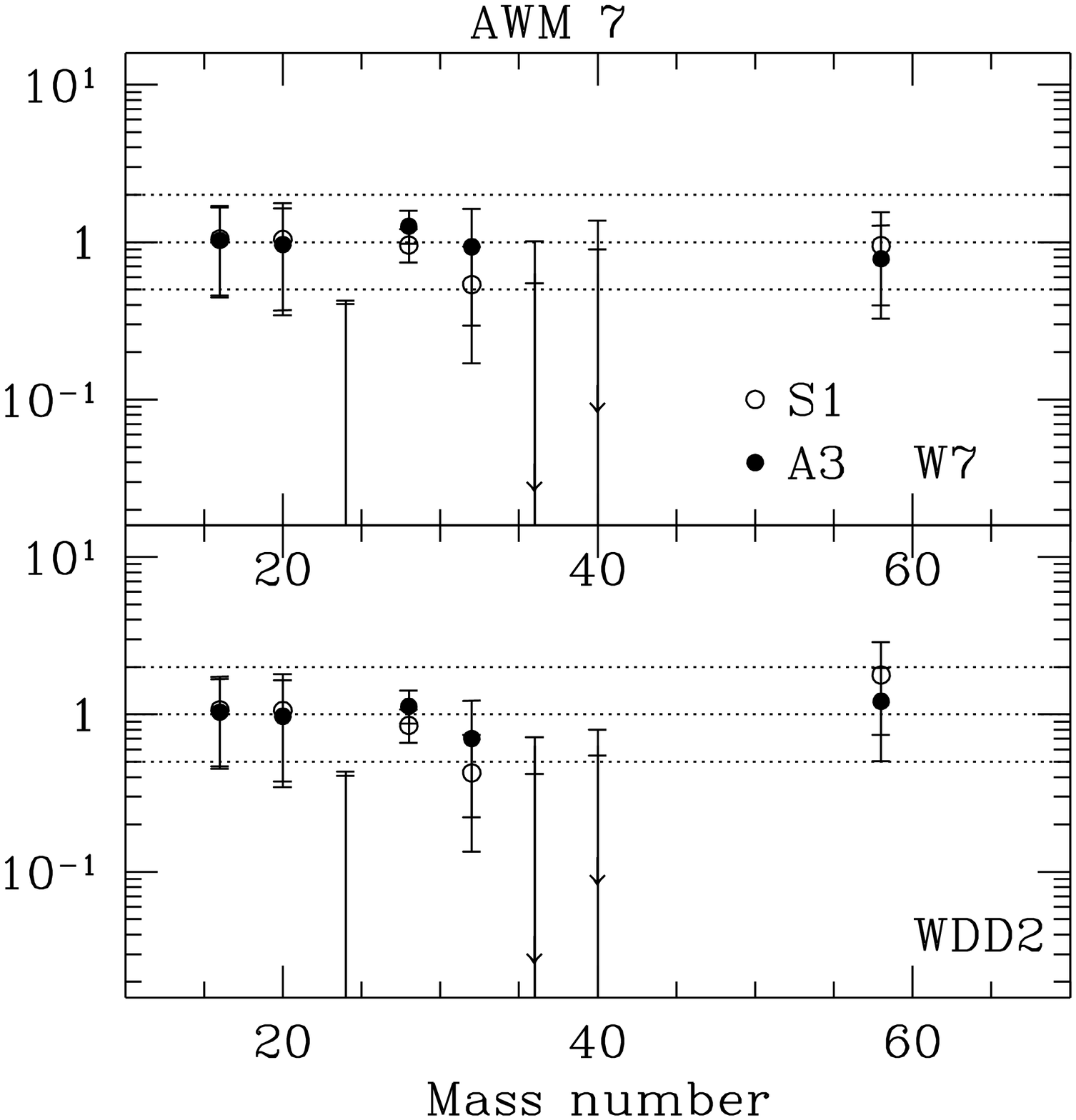}
\figcaption{
Same with figure 7 but for Abell 2199 and AWM 7. 
\label{fig8}}
\end{figure}

Finally, Figure ~\ref{fig7} and ~\ref{fig8} show the normalized
abundance pattern $(x_i/x_{\rm Fe})_{\rm ICM} / (\it
x_i/x_{\rm Fe})_{\rm SN}$ with the most probable value $\zeta =
\zeta_p$ which is obtained from the analysis of $g(\zeta)$ for each
cluster.
The abundance of sulfur seems to be reproduced better by N97A3 
than N97S1, to be sure, however, the sulfur abundance can be
reproduced within a factor of 2 even by N97S1. We note that the
supernova abundances agree with the solar system abundance ratios
within a factor of 2-3 for typical species (\cite{hashimoto95}), which 
will reflect the precision of the numerical calculations of supernova
nucleosynthesis. Because of this, we think the sulfur abundance 
problem needs not to be emphasized too strongly. We wait for
more precise observations to conclude whether ICMs really suffer the
shortage problem of sulfur or not.

\placefigure{fig7}
\placefigure{fig8}

\section{ Summary and Discussion} \label{summary}
\indent

In this paper, the origin of the chemical composition in ICMs is
discussed. At first, the analysis which has been done in the previous papers
is reexamined. As expected, the ratio of $M_{\rm Fe, SNIa}/M_{\rm Fe,
Total}$ is very hard to determine. 
Quantitatively, the observations of O, Ne, Si, S, and Fe favor the
ICM SNe Ia iron fractional contribution in the range $30 \% \le
M_{\rm Fe, SNIa}/M_{\rm Fe, Total} \le 60 \%$ for the models W7 and N97A3.
The range is $30 \% \le M_{\rm Fe,
SNIa}/M_{\rm Fe, Total} \le 90 \%$ for the models W7 and N97S1.
Even worse, the averaged [S/Fe] ratio can not be reproduced
when model WDD2 is adopted.
Our analysis shows that $M_{\rm Fe, SNIa}/M_{\rm Fe,
Total}$ is not a good measure for estimating the
contribution of SNe Ia at the moment to the ICMs.

Next, the analysis of $g( \zeta )$ and $\chi ^2$ fitting are done in
order to investigate whether these can be better measures for SNe Ia's 
contribution. We have shown that $g(\zeta)$ and $\chi
^2$ have a local minimum/maximum for
each model, which means there is a favorable value of $\zeta$ for the
reproduction of the chemical composition of ICMs.
Moreover, $\zeta_p$ is in the range of $0 < \zeta_p  < 0.1$ for almost all
clusters, which means that the contributions of SNe Ia to the ICM
enrichment are similar to each other and to the solar system abundances.
This analysis indicates that $\zeta$ is a better measure
than $M_{\rm Fe, SN Ia}/M_{\rm Fe, total}$.

The reason why $\zeta_p$ can be determined in the narrow
range $\zeta_p \le 0.1$ in spite of the large uncertainty of the
inferred ratio $M_{\rm Fe, SN Ia}/M_{\rm Fe, total}$ is that $M_{\rm
Fe, SN Ia}/M_{\rm Fe, total}$ is very sensitive to $\zeta$
if $\zeta$ is in the range of $[0,0.1]$.
This is because the mass fraction of Fe in the SNe Ia's ejecta is quite high.

Finally, we have investigated the sulfur abundance problem.
The normalized abundance pattern $(x_i/x_{\rm Fe})_{\rm ICM} / (\it
x_i/x_{\rm Fe})_{\rm SN}$ with the most probable value $\zeta =
\zeta_p$ which is obtained from the analysis of $g(\zeta)$ shows that
the sulfur abundance can be reproduced within a factor of 2.
Taking the uncertainty of the numerical calculation of supernova
nucleosynthesis, we think that the sulfur abundance 
problem needs not to be emphasized too strongly. We need
more precise observations to conclude whether ICMs really suffer the
shortage problem of sulfur or not. The same applies for Ar and Ca.
We hope that there will be a large number of more precise observations 
of ICMs to have a more advanced discussion.

\acknowledgements
We would like to thank W. Hillebrandt for useful comments.
We are also grateful to V. Saar for his
kind review of the manuscript.
This research has been
supported in part by a Grant-in-Aid for the Center-of-Excellence (COE) 
Research (07CE2002) and for the Scientific Research Fund (05243103,
07640386, 3730) of the Ministry of Education, Science, and Culture in Japan
and by Japan Society for the Promotion of Science Postdoctoral
Fellowships for Research Abroad.

\clearpage

\begin{table*}
\small
\begin{center}
\begin{tabular}{lccccccccccccccccccccc}
\tableline
\tableline
        & Abell 496 &  Abell 1060  &  Abell 2199 & AWM 7 \\
\tableline
$\rm [O/Fe]$  & $+0.15(-0.15,+0.36)$ & $-0.08(-0.33,+0.08)$   &
$-0.06(-0.45,+0.15)$& $-0.01(-0.37,+0.20)$                   \\
$\rm [Ne/Fe]$ & $+0.28(+0.10,+0.43)$ & $+0.15(+0.00,+0.25)$   &
$+0.12(-0.12,+0.28)$& $-0.16(-0.61,+0.07)$                   \\
$\rm [Mg/Fe]$ & $-0.01(-0.43,+0.21)$ & $-0.31(-0.92,-0.06)$   &
$-0.05(-0.48,+0.18)$& -----  $(<-0.39)^{\dagger}$                   \\
$\rm [Si/Fe]$ & $+0.06(-0.09,+0.18)$ & $+0.14(+0.07,+0.21)$   &
$+0.26(+0.16,+0.33)$& $+0.03(-0.08,+0.13)$                   \\
$\rm [S/Fe]$  & $-0.23(-0.62,-0.02)$ & $-0.30(-0.55,-0.15)$   &
$-0.31(-0.86,+0.28)$& $-0.47(-0.97,-0.23)$                   \\
$\rm [Ar^{\dagger}/Fe]$ & $-0.18(<+0.20)$      & $-1.31(<-0.23)$        &
-----  $(<-0.31)$ & -----  $(<-0.52)$                  \\
$\rm [Ca^{\dagger}/Fe]$ & $-0.25(<+0.20)$      & -----  $(<-0.23)$        &
-----  $(<-0.09)$ & -----  $(<-0.39)$                   \\
$\rm [Ni/Fe]$ & $+0.49(+0.25,+0.65)$ & $+0.23(-0.33,+0.61)$   &
$+0.37(+0.00,+0.54)$& $+0.19(-0.19,+0.40)$        \\
\tableline
\end{tabular}
\end{center}

\tablenum{1}
\caption{
Abundances of ICM normalized by the meteoritic abundances. $\dagger$ :
Nuclei omitted in the analysis of $g(\zeta)$.
\label{obs}}

\end{table*}

\begin{table*}
\small
\begin{center}
\begin{tabular}{lccccccccccccccccccccc}
\tableline
\tableline
Yield Source& $<y_{\rm O,SNII}>$ & $<y_{\rm Ne,SNII}>$  &  
$<y_{\rm Mg,SNII}>$  & $<y_{\rm Si,SNII}>$ & $<y_{\rm S,SNII}>$ & $<y_{\rm Fe,SNII}>$ \\
\tableline
A96                       & 0.593   & 0.101 & 0.054 & n/a   & n/a   & 0.071 \\
T95                       & 1.777   & 0.232 & 0.118 & 0.133 & 0.040 & 0.121 \\
T95+M92                   & 0.923   & n/a   & n/a   & n/a   & n/a   & 0.121 \\
W95;A;$10^{-4}Z_{\odot}$  & 0.806   & 0.095 & 0.036 & 0.104 & 0.059 & 0.073 \\
W95;B;$10^{-4}_{\odot}$   & 1.455   & 0.223 & 0.066 & 0.118 & 0.065 & 0.085 \\
W95;A;$Z_{\odot}$         & 1.217   & 0.181 & 0.065 & 0.124 & 0.058 & 0.113 \\
W95;B;$Z_{\odot}$         & 1.664   & 0.265 & 0.094 & 0.143 & 0.064 & 0.141 \\
N97S1                      & 1.749   & 0.229 & 0.124 & 0.139 & 0.042 & 0.086 \\
N97A1                      & 1.726   & 0.231 & 0.126 & 0.106 & 0.026 & 0.086 \\
N97A3                      & 1.694   & 0.233 & 0.123 & 0.098 & 0.020 & 0.088 \\
\tableline
Yield Source& $<y_{\rm O,SNIa}>$ & $<y_{\rm Ne,SNIa}>$  &  
$<y_{\rm Mg,SNIa}>$  & $<y_{\rm Si,SNIa}>$ & $<y_{\rm S,SNIa}>$
& $<y_{\rm Fe,SNIa}>$ \\
\tableline
W7                        & 0.148   & 0.005 & 0.009 & 0.158   & 0.086  & 0.744 \\
WDD2                      & 0.069   & 0.0009 & 0.005 & 0.272  & 0.168   & 0.700 \\
\tableline
\end{tabular}
\end{center}

\tablenum{2}
\caption{
Average stellar yield in solar masses for the SNe grids.
A96 = Arnett 1996, T95 = Tsujimoto et al. 1995, M92 = Maeder 1992, W95 
= Woosley $\&$ Weaver 1995, N97 = Nagataki et al. 1997c. N97S1, N97A1, and
N97A3 stand for the model of the spherical explosion and those of the
axisymmetric explosion, respectively. The degree of deviation from
spherical symmetry is
larger in N97A3 than N97A1.
W7 is the model of the simple deflagration and WDD2 is that of the
delayed detonation (Nomoto et al. (1997)).
\label{sneii}}

\end{table*}

\end{document}